\documentclass[twocolumn,superscriptaddress, aps,pra ]{revtex4-1}
\makeatletter
\newcommand*{\rom}[1]{\expandafter\@slowromancap\romannumeral #1@}
\makeatother
\usepackage{color}
\usepackage{graphicx}
\usepackage{hyperref}
\usepackage{amsmath,amssymb}
\usepackage{epstopdf}
\usepackage{subcaption}
\graphicspath{{figs/}}
\def\beq{\begin{equation}}
\def\eeq{\end{equation}}
\def\bea{\begin{eqnarray}}
\def\eea{\end{eqnarray}}

\begin{document}

\title{Deformed Boson Condensate as a Model of Dark Matter}
\author {Mahnaz Maleki}
\email{m.maleki@uma.ac.ir}
\affiliation{Department of Physics, University of Mohaghegh Ardabili, P.O. Box 179, Ardabil, Iran}
\author {Hosein Mohammadzadeh}
\email{mohammadzadeh@uma.ac.ir}
\affiliation{Department of Physics, University of Mohaghegh Ardabili, P.O. Box 179, Ardabil, Iran}
\author {Zahra Ebadi}
\affiliation{Department of Physics, University of Mohaghegh Ardabili, P.O. Box 179, Ardabil, Iran}
\author {Morteza Nattagh Najafi}
\affiliation{Department of Physics, University of Mohaghegh Ardabili, P.O. Box 179, Ardabil, Iran}
\begin{abstract}
We consider the condensate of $q$-deformed bosons as a model of dark matter.
Our observations demonstrate that for all $q$ values, the system condenses below a $q$-dependent critical temperature $T^{q}_c$. The critical temperature interestingly tends to infinity when $q\rightarrow 0$, so that the $q$ deformed boson gas is always in the condensed phase in this limit irrespective to the temperature.
We argue that this has remarkable outcomes, e.g. on the entropy of the system, and also the fraction of the particles in the ground state. Especially, by direct evaluation of the entropy of the system we reveal that it tends to zero at this limit for all temperatures, and also the fraction of particles in the ground state becomes unity. These observations prove the consistency of the model, put it in the list of appropriate candidates for the dark matter. Also, the lower and upper bounds of mass are evaluated using the phase space density and observational data for $q$ deformed Bose-Einstein condensate ($q$-BEC) model.
\end{abstract}
\maketitle

\section{Introduction}\label{1}
One half of the Nobel prize in physics 2019 was awarded to Phillip James Edwin Peebles for contributions to understanding of the evolution of the universe and for theoretical discoveries in physical cosmology.
These works involve the Big Bang nucleosynthesis, dark matter, and dark energy and the cosmic structure formation \cite{peebles2015physical,peebles1980large}. In fact, the standard model of cosmology is indebted to his efforts.

Although such studies have opened new avenues in the physical cosmology, we are suffering yet a big lack of knowledge in this area.
The standard model of cosmology suggests that approximately $4$\% of the total energy content of the universe is made up of ordinary baryonic matter, and the remaining $96$\% is composed of an unknown form of matter or energy, that is, $22$\% of this content is an unknown matter that it is a non-baryonic matter called dark matter, and $74$\% of the remaining content is composed of unknown energy called dark energy\cite{weinberg2008cosmology,jarosik2011seven,lyth2009primordial}.
One of the important issues in the standard model of cosmology is about the nature of dark matter.

It is known that to explain the velocity of the galaxies in a cosmic cluster, a missing mass is required other than the luminous matter. Also, the cosmology nucleosynthesis investigation and observed anisotropies in Cosmic Microwave Background (CMB) confirms the existence of an unknown matter in galaxy and cosmic clusters formation and consequently affects the motion of galaxies\cite{weinberg2008cosmology}. By studying the observational data about the galaxies rotation profile and gravitational lensing, it is concluded that an important part of the galaxies disk is made up of dark matter \cite{trimble1987existence,durrer1997cosmic,nesti2013dark,gorenstein2014astronomical}. For explaining these cosmological observations, various models for the dark part of the universe were proposed. One of these models is called ($\mathrm{\Lambda CDM})$ which is based on the assumption that the particles have no charge, and are cold and long lived \cite{weinberg2008cosmology}. Also, the astrophysical observations demonstrate that the dark matter is composed of Weakly Interacting Massive Particles (WIMP). In fact, the theoretical considerations imply that the particles of the dark matter should be discovered in beyond the baryonic matter, such as supersymmetric particles or extra high energetic particles. Although the direct detection of dark matter seems not to be possible, its detection indirectly via annihilation (to standard matter in the sun, or in the experiments in the large Hadron collider (LHC)), and upcoming electron-positron linear collider is possible \cite{feng2010dark,raffelt1996stars}.

Another popular model for the dark matter is Bose-Einstein condensate (BEC). The thought that dark matter is in the form of BEC was proposed initially in \cite{membrado1989statistical,ji1994late}, and restudied in seminal works \cite{guzman2000scalar,lee1996galactic,lee2008dark,kain2012cosmological,urena2009bose,boehmer2007can,das2015dark}. At very low temperature, all particles of an ideal Bose gas condense into the same quantum ground state, making a BEC. The BEC transition takes place at a temperature $T^{b}_{c}$ under which a considerable fraction of particles condense into the ground state \cite{pitaevskii2016bose,pethick2008bose,griffin2009bose}. Different properties of the BEC model of dark matter have been investigated recently in \cite{fukuyama2008cosmic,fukuyama2009stagflation,brook2009gravitational,kain2010vortices}.
Using the Gross-Pitaevskii equation, some characteristics of dark matter such as density, velocity of rotation and mass profile is considered \cite{harko2011cosmological,harko2011bose,harko2015testing,zhang2018slowly}. Of course, BEC has been utilized in some other aspects of gravitation and cosmology \cite{zare2012condensation,raffaelli2013scattering}.

In the BEC model of dark matter, boson particles are considered as the main component of dark matter and the Bose-Einstein condensation occurs at low temperature. {One may assume that the temperature
of the dark matter candidate is equal to the temperature of the Universe. Nowadays, the average temperature of the universe is low and about 2.7 k. If Big Bang model is a valid model for the creation of the universe, it is expected that in the far past epochs, the temperature of the universe should be more than the current situation. Therefore, the models describing dark matter based on BEC, lose their reliability. In fact, the BEC model of dark matter can not be a valid model in all epochs. However BEC dark
matter are typically composed of light particles, and they are necessarily
produced by non thermal mechanisms. This is due to the fact that if such
light particles are produced by thermal processes, then it will result in a too
large number of relativistic degrees of freedom \cite{boyanovsky1999nonequilibrium,earlykolb1990,baumann2014lecture}.
In this case, the temperature of the dark matter candidate is not equal to
the age of the Universe.}  The situation gets better when one considers the particles with deformed statistics, exploiting of which one can have a generalized condensate in all epochs. This is the aim of the present paper. Using deformed statistics, we will introduce a generalized statistics condensate as a good model of dark matter.

There are different generalized statistics, each of which has different origins. Infinite statistic, $q$-bosons and $q$-fermion, $\mu$-bosons, fractional exclusion statistics and non-extensive statistics are examples of such statistics \cite{greenberg1990example,mirza2010thermodynamic,mirza2011thermodynamic,gavrilik2011intercepts,mohammadzadeh2017thermodynamic,mohammadzadeh2016perturbative}. Recently, it has been shown that  condensate of infinite statistic can be a good alternative to the particles of dark matter\cite{ebadi2013infinite}. Also, condensate of $\mu$-Bose gas as a model of dark matter has been proposed \cite{gavrilik2018condensate} and more recently the galaxy rotation curves has been considered based on the $\mu$-deformation approach \cite{gavrilik2019galaxy}. Also, the effective dark matter theory has been studied in electron-positron annihilation in a typical supernova explosion using deformed statistics \cite{guha2017q}.

The rest of the paper is organized as follows. In section \ref{sec2}, we review the $q$-deformed algebra. We derive the internal energy and particle number of $q$-deformed statistics. In section \ref{sec3}, we obtain the critical condensation temperature of $q$-bosons. Also, we consider the $q$-boson condensate as an effective model for dark matter and show that for small values of deformation parameter, it can be a valid model for all epochs due to the very high critical condensation temperature in section \ref{sec4}. Also, we study the equation of state and entropy of $q$-boson condensate and show that the results are compatible with properties of dark matter. In section \ref{sec5}, we consider the relic density and mass bounds of the model. Finally, we conclude the paper in section \ref{sec6}.
\section{$q$-deformed statistics}\label{sec2}
In this section we briefly review the $q$-deformed statistics concentrating mainly on the average occupation number.
The symmetric $q$-oscillator algebra is defined in terms of the creation operator ($a^\dag$), annihilation operator ($a$) and the $q$-number operator ($N=a^\dag a$) as follows\cite{biedenharn1989quantum,macfarlane1989q,tuszynski1993statistical}
 \bea
aa^{\dag}-kqa^{\dag}a=q^{-N},~~~~~~ [a,a]_{k}=[a^{\dag},a^{\dag}]_{k}=0,
\label{a}
 \eea
 \bea
[N,a^{\dag}]=a^{\dag},~~~~~~[N,a]=-a,
 \eea
 \bea
a^{\dag}a=[N],~~~~~~~~~~~aa^{\dag}=[1+k N],
\label{b}
 \eea
where $[a,b]_k=ab-kba$ and $k=1 (k=-1)$ is for $q$-bosons ($q$-fermions). Working in the $q$-based numbers facilitates much the calculations in this algebra. The numbers in base of $q$ are written as
\bea
[x]=\frac{q^{x}-q^{-x}}{q-q^{-1}}.
\label{c}
\eea
Also, the Jackson derivative (JD) is of great help in the following calculations for $q$-algebra, defined as follows
\bea
{\cal{D}}^{(q)}_{x}f(x)=\frac{f(q x)-f(q^{-1}x)}{(q-q^{-1})x}.
\label{d}
\eea
Now, we consider the following Hamiltonian of non-interacting $q$-deformed oscillators
\bea
 H=\sum_{i}(\epsilon_{i}-\mu)N_{i},
\label{e}
\eea
where, the index $i$ is the state label, $\mu$ is the chemical potential and $\epsilon_{i}$ is the kinetic energy in the state $i$ with the number operator $N_i$. We can define $q$-basic mean occupation number by
\bea
[n_i]=\frac{\mathrm{Tr}(e^{-\beta H}a^{\dag}_{i}a_{i})}{{\mathrm{Tr}(e^{-\beta H})}},
\label{f}
\eea
 where, $\beta=\frac{1}{k_BT}$ and $k_B$ is the Boltzmann constant. It is straightforward to obtain the mean value of the occupation number or distribution function as follows \cite{mirza2010thermodynamic,mirza2011thermodynamic}
\bea
{n}_{i}= {f(\epsilon )}=\frac{1}{q-q^{-1}}\ln\left(\frac{z^{-1}e^{\beta\epsilon_{i}}-kq^{-k}}{z^{-1}e^{\beta\epsilon_{i}}-kq^{k}}\right),
\label{h}
\eea
where, $z=e^{\beta \mu}$ is the fugacity. Therefore, the internal energy and particle number of an ideal gas with $q$-deformed statistic in a $d$ dimensional box are given by \cite{mirza2011thermodynamic}
\bea
 U&=&\int_{0}^{\infty}\epsilon
 {f(\epsilon )}\Omega(\epsilon )d\epsilon  \nonumber\\
&=&\frac{d}{2}\frac{V}{(q-q^{-1})}(\frac{2\pi mk_{B}T}{h^{2}})^{d/2} (k_{B}T){H}_{d/2+2} ( z,k,q),~~~\\
 {N}&=&\int_{0}^{\infty}{f(\epsilon )} \Omega(\epsilon)d\epsilon \nonumber\\
&=&\frac{V}{(q-q^{-1})}(\frac{2\pi mk_{B}T}{h^{2}})^{d/2}{H}_{d/2+1}( z,k,q),
\label{i}
\eea
where,
\bea
 {H}_{n}(az,k,q)=Li_{n}(azkq^{k})-Li_{n}(azkq^{-k}),
\label{n}
\eea
and $Li_{n}(x)$ denotes the polylogarithm function. We note that the density of the single particle state is
\bea
\Omega(\epsilon)=\frac{V}{\Gamma(\frac{d}{2})}\frac{(2m\pi)^{d/2}}{h^{d}}\epsilon^{d/2 -1},
\label{j}
\eea
where $V$ is the volume of the system, $m$ is the mass of particle, $\Gamma(n)$ is the well known gamma function and $h$ is the planck constant. Also, we suppose that the dispersion relation is $\epsilon=p^{2}/2m$.
\section{$q$-boson condensation}\label{sec3}
\begin{figure}
\includegraphics[scale=0.45]{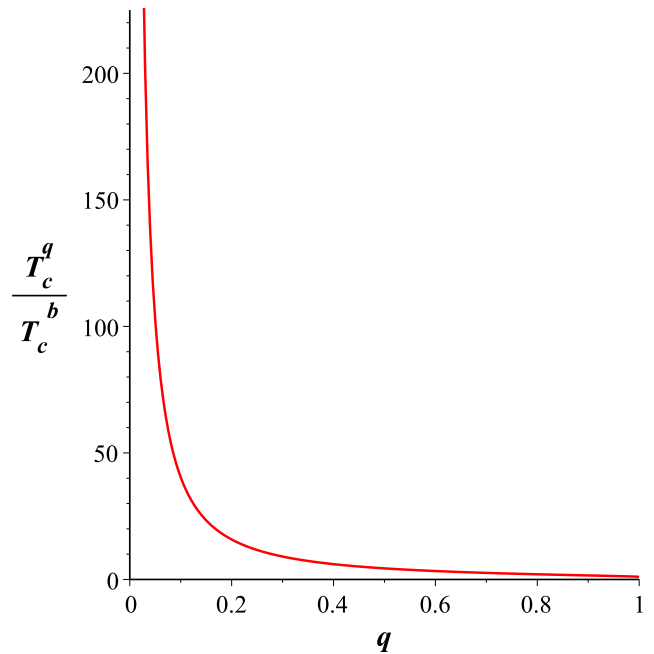}
\caption{the ratio of the critical condensation temperature  of deformed bosons as a function of the deformation parameter.}
\label{fig1}
\end{figure}

Thermodynamic behaviours of $q$-bosons and $q$- fermion vastly has been investigated by several authors \cite{lavagno2000thermostatistics,lavagno2002generalized,lavagno2010thermostatistics,mirza2011thermodynamic}. Using thermodynamic geometry method, it has been shown that the intrinsic statistical interaction of an ideal gas with $q$-bosons is attractive. Also, it has been shown that the thermodynamic curvature is singular at a critical value of fagucity which is given by \cite{mirza2011thermodynamic}
\bea
 z_q=\left\{
         \begin{array}{cc}
           q^{2} & q<1 \\
           q^{-2} & q>1 \\
         \end{array}\right.\label{zq}.
\label{k}
\eea
It is well-known that the thermodynamic curvature of ordinary boson gas is singular at $z=1$, where the Bose-Einstein condensation (BEC) occurs. Similar to BEC, the critical fugacity could be related to the condensation of $q$-deformed boson gas. We restrict ourselves to the range that the deformation parameter belongs to $0\le q\le 1$. In fact, in the limit of $q\rightarrow 1$, the problem reduces to the ordinary bosons and the small values of deformation parameter correspond to the more deformed cases. Using the particle number find the conditions under which the $q$-boson system condensates. Interestingly we found that for all considered $q$ values, the condensation occur under a critical temperature $(T_c^q)$ satisfying the relation
\bea
n=\frac{N}{V}&=&\frac{1}{(q-q^{-1})}\left(\frac{2\pi m_q k_{B}T_{c}^q}{h^{2}}\right)^{d/2}{H}_{d/2+1}(q^{2},1,q),~~~~
\label{l}
\eea
from which we obtain
\bea
  k_{B}T_{c}^q=\frac{2\pi\hbar^{2}}{m_{q}}\left[\frac{n(q-q^{-1})}{H_{\frac{d}{2}+1}(q^{2},1,q)}\right]^{2/d},
\label{TCQ}
\eea
where, $m_q$ refers to the mass of $q$-bosons.
Also, for an ideal boson gas, the critical temperature of  Bose-Einstein condensation is evaluated as follows \cite{pathria1996statistical}
\bea
 k_{B}T_{c}^b=\frac{2\pi \hbar^{2}}{m_{b}}(\frac{n}{\zeta(d/2)})^{2/d}.
\label{TCB}
\eea
where, $m_{b}$ is the mass of bosons and $\zeta(x)$ is the Riemann zeta function. Comparing the ordinary BEC and $q$-deformed condensates, we deduce some interesting results. We evaluate the ratio of critical condensation temperature of $q$-bosons to condensation temperature of ordinary bosons as follows
\bea
 \frac{T_{c}^q}{T_{c}^b}=\frac{m_{b}}{m_{q}}\left(\frac{(q-q^{-1})\zeta(d/2)}{H_{d/2+1}(q^{2},1,q)}\right)^{2/d}.
\label{TCQB}
\eea
 Eq. (\ref{TCQB})  shows that the critical temperature depends on the value of deformation parameter.We suppose that $m_{q}=m_{b}$ and $d=3$ and depict the critical condensation  temperature as a function of deformation parameter in Fig. (\ref{fig1}).
\\
 It is obvious that the critical temperature of $q$-bosons goes to infinity for enough small values of deformation parameter, while it tends to the critical temperature of ordinary bosons $(T^{b}_{c})$ at the limit $ q\rightarrow 1$. We can argue that $q$-deformed boson gas with enough small deformation parameter will be in condensate phase at any finite temperature because of the large value of critical temperature.

Another view point about the $q$-bosons seems sound and interesting. Eq. (\ref{TCB}) indicates that the heavier the bosons are, the lower the critical condensation temperature is. In fact, by increasing the mass of bosons, the condensation temperature tends to the zero temperature. Therefore, possibility of condensation of heavy bosons is less than the light ones.  However, using Eq. (\ref{TCQB}) we can argue that the at the same value of critical temperature for ordinary and $q$-boson condensation, more heavier deformed bosons in comparison with ordinary bosons could exist in condensed phase, specially for small value of deformation parameter.

Up to the authors' knowledge, this is the first observation of such a phenomenon, i.e. the infinite condensation critical temperature. As we will see in the following sections, this opens new possibilities for explaining the less-known features of the dark matter.
\section{$q$-deformed condensate and dark matter}\label{sec4}
Recently, it has been proposed that BEC could be an appropriate candidate for describing the dark matter. Some useful information about the velocity profile of different galaxies is extracted by BEC model \cite{harko2011cosmological,harko2011bose,harko2015testing}. Also, condensate of infinite statistics as a candidate for dark matter has been investigated\cite{ebadi2013infinite}.
An important question exists about the BEC models for dark matter. Are such models valid for all epochs?
Once such models are built, the important question arises concerning the time evolution of the condensates. By this we don't mean the time dependent evaporation of BEC, but the time evolution of the equation of state of the condensate. This evolution contains, but is not restricted only to the thermodynamic quantities,
like the temperature and the volume etc., but involves the time evolution of the intrinsic parameters like $q$. As outlined in Ref. \cite{algin2017effective,mohammadzadeh2017thermodynamic}, the deformation parameters can be related to the interaction between ordinary Bosons, so that by the expansion of the system one expects that this quantity is variable. At the moment, there is no equation describing the time evolution of $q$, but by some general considerations, some information can be obtained, which is the aim of the present paper.

Although, the current effective temperature of universe is low, we expect a high temperature for early universe. Therefore, condensate of boson gas as a candidate for dark matter fails down for past epochs, more specially for early universe. In other words, since the temperature of the early universe was very high, BEC could not be stable for that stage, and the BEC theory of Ref. \cite{harko2011cosmological,harko2011bose,harko2015testing} fails for hot universe. Also if the dark matter has minimal interaction with luminous matter, which means that it is nearly conservative the problem of entropy of dark matter remains unanswered: what happens to the lost entropy, when the dark matter condensates at the time that the temperature of the universe was suitable to BEC?

We considered the condensation of a q-deformed boson gas in previous section. We showed that the critical condensation temperature becomes very high for a deformed bosons with small deformation parameter. In fact, the critical temperature goes to infinity at the limit of $q \rightarrow 0$. Therefore, a deformed gas with small deformation parameter will be in condensate phase and can be a suitable model for dark matter in all time durations.

Recently, it has been shown that if dark matter is assumed to consist of an ideal gas of ordinary bosons of mass $m_b$, then for $m_{b}\le 1 eV$, the critical condensation temperature below which they will form a Bose–Einstein condensate exceeds the temperature of the universe at all times \cite{das2015dark}. According to the arguments of last paragraph of pervious section, there are no restriction on the mass of deformed bosons to form a condensate at all times for enough small values of deformation parameter. Our model, in addition to relaxing this condition, has other strengths and benefits that are explained in the following.
\subsection{Equation of state in condensed phase}
Experimental results and theoretical arguments suggest that the interaction between particles of dark matter should be very weak, so we expect that the equation of state of it is just the same as the ideal gas. Therefore, in this subsection we show that our model satisfies this, i.e. the equation of state of the condensate of $q$-bosons is of the same for as the ideal gas, and then in the next subsection we turn to the key observations and explanations. By a simple evaluation, we can show that 
\bea
PV=\frac{2}{3}U,
\label{t1}
\eea
and using Eq.(9), we work out that in condensed phase the equation of states reads
\bea
PV=\gamma Nk_{B}T,
\label{u}
\eea
where,
\bea
\gamma=\frac{{H}_{\frac{7}{2}}(q^{2},1,q)}{{{H}_{\frac{5}{2}}(q^{2},1,q)}}.
\label{w}
\eea
\begin{figure}[htp]
\includegraphics[scale=0.45]{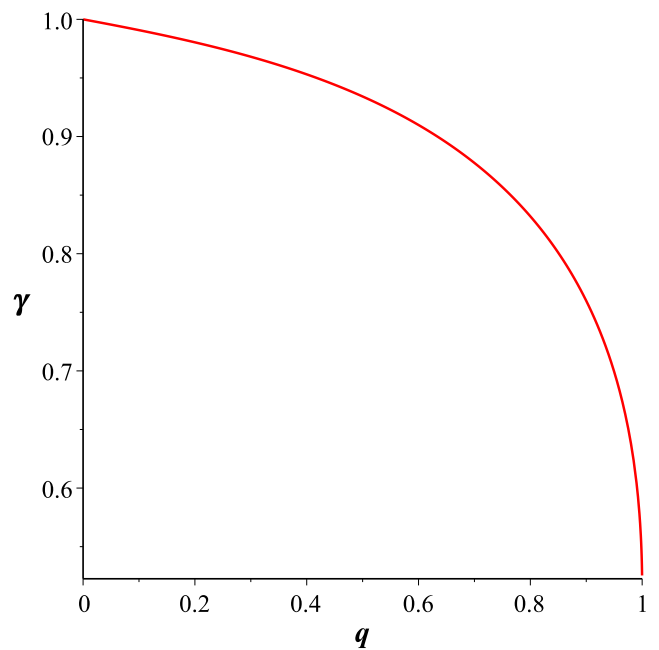}
\caption{The $\gamma$ coefficient with respect to the deformation parameter.}
\label{fig2}
\end{figure}
We plot the coefficient $\gamma$ as a function of deformation parameter in Fig. (\ref{fig2}). It is obvious that that at the limit of small values of deformation parameter ($q\rightarrow0$), the coefficient $\gamma\rightarrow1$. Therefore, the equation of state of an ideal gas with $q$-deformed bosons becomes the equation of state of an ideal classical gas without any interaction, $PV= Nk_{B}T$ for small $q$s, justifying that the condensate of $q$-bosons is compatible with non-interacting dark matter.
\subsection{Entropy of $q$-bosons condensate}
We can assign some problems about the entropy of ordinary Boltzmann and Bose-Einstein statistics, if constituents of dark matter obey such statistics. For an ideal classical gas in a box with volume $V$, we know that the partition function is
\bea
Z_{N}=\frac{1}{N!}(\frac{V}{\lambda^{3}})^{N},
\label{x1}
\eea
where $\lambda=h/\sqrt{2m\pi k_{B}T}$ is the thermal wavelength\cite{pathria1996statistical}. Also, we calculate the entropy of the system using the free energy which is given by $A=-k_{B}T\ln Z_{N}$, and obtain that
\bea
S=-(\frac{\partial A}{\partial T})_{V,N}=Nk_{B}\left(\ln(\frac{V}{N\lambda^{3}})+\frac{5}{2}\right).
\label{x2}
\eea
It is obvious that by decreasing the temperature, the thermal wavelength grows up. Clearly at a certain sufficiently low temperature  $V\sim\lambda^{3}$ while $N\gg1$ and consequently the entropy of the system will become negative\cite{ng2007holographic}. In fact, the origin of this problem is related to the Gibbs $1/N!$ factor due to the distinguishability of the classical particles.
For an ideal boson gas in condensate regime, evaluation of the entropy is straightforward and is given by
\bea
S^{b}= k_{B}\frac{5}{2}\frac{V}{\lambda^3}\zeta(\frac{5}{2}), ~~~~T\le T^{b}_{c}.
\eea
for which the negative entropy problem does not further exist. However, the entropy of the system depends on the temperature. Decreasing the temperature causes growing up the thermal wavelength and consequently the entropy of the system is reduced. BEC model of dark matter in a typical galaxy, produces another problem: the entropy of the system gradually decreases from the early universe to current epoch. Compensation of this reduction by the environment around (luminous matter) needs some interactions between dark matter and other constituents of universe, in contrast with the hypothesis that the dark matter should be non interacting.

We show that the entropy of the condensate of a $q$-deformed boson gas,  has none of the problems mentioned above. Especially we argue that the entropy is not negative for deformed bosons for any temperature. To find that out, by introducing the Helmholtz free energy $A=\mu N-PV$, we represent the entropy as follows
\bea
S=\frac{U-A}{T}=\frac{U+PV-\mu N}{T},
\label{x4}
\eea
where $\mu$ is the chemical potential
\bea
\mu=k_{B}T \ln z.
\label{x5}
\eea
In the condensate phase, the fugacity is shown to be $z_q=q^2$, leading $\mu$ to have the form:
\bea
\mu=k_{B}T \ln q^{2},~~~~~T\le T^{q}_{c}.
\label{x6}
\eea
Using Eqs. (\ref{i}), (\ref{t1}) and (\ref{x6}) and incorporating them into Eq. (\ref{x4}), one obtains the entropy of $q$-deformed boson gas in the condensate phase as follows
\bea
S^{q}=\frac{k_{B}}{q-q^{-1}}\frac{V}{\lambda^{3}}\left(\frac{5}{2}{H}_{7/2}(q^{2},1,q)-{H}_{5/2}(q^{2},1,q)\ln(q^{2})\right).\nonumber\\
\label{x7}
\eea
\begin{figure}[htp]
\includegraphics[scale=0.45]{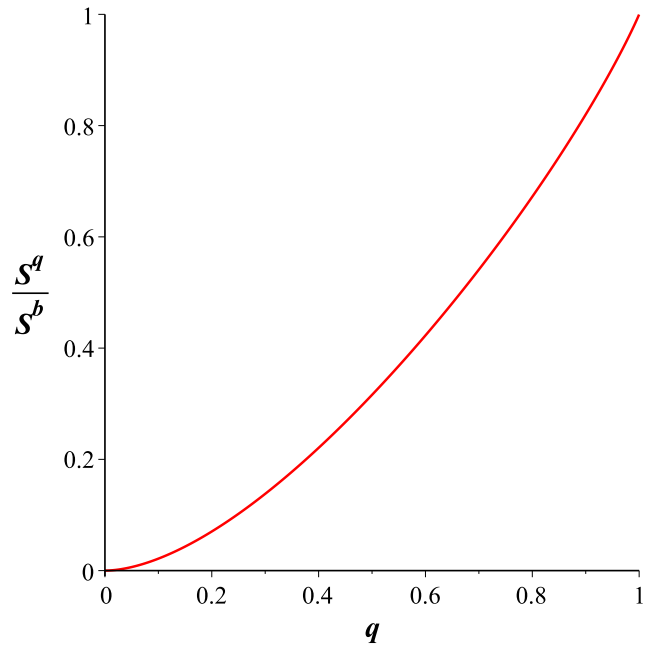}
\caption{The entropy; $S$, of deformed boson condensate with respect to the deformation parameter.}
\label{fig3}
\end{figure}
It is clear in Fig. (\ref{fig3}) that the entropy of deformed condensate is equal to the entropy of boson condensate at the limit of $q=1$. However, it is interesting that the entropy of condensate of $q$-deformed bosons is vanished at the limit of $q\rightarrow 0$. Of course, the zero entropy of $q$-boson condensate occurs at any finite temperature. We argued that in the limit of small values of deformation parameter, the critical condensation temperature goes to infinity and the system lives in condensate regime at any finite temperature and the entropy of the system is zero. Therefore, variation of temperature in different time duration, does not affect the entropy of the system. Therefore, $q$-boson condensate as a model of dark matter does not need to interact with surrounding or remnant of the universe to compensate the reduction of entropy.

\begin{figure}[htp]
\includegraphics[scale=0.45]{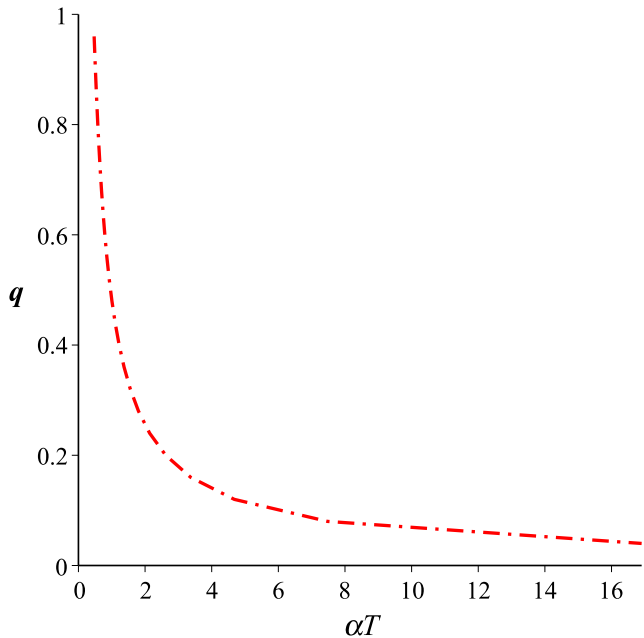}
\caption{Variation of deformation parameter with respect to the temperature of each epoch. $\alpha$ is a constant which depends on the value of physical constants such as $m$, $\hbar$, $k_{B}$, $V$ and fixed value of entropy. $T$ is the absolute temperature.}
\label{fig4}
\end{figure}

Another viewpoint to the entropy reduction problems seems interesting. We could interpret the deformation parameter as a dynamical parameter which varies with temperature of each epoch, while the entropy of the system remains fixed. Fig. (\ref{fig4}) shows the general behavior  of deformation parameter with temperature of each epoch. It is obvious that at early universe which we expect very high temperature, the deformation parameter is small. As time goes on and the temperature decreases, the deformation parameter grows up and tends to unity and the statistics of particles turns to the ordinary bosons while the entropy of the system is fixed.

Here we propose two main scenarios. The first scenario is that $q$ is fixed for all times (during the expansion of the universe), and it is small enough that for all stages the BEC phase is stable, so that its entropy is fixed and no other scenarios are needed to explain the entropy reduction.
The second scenario, is the time-dependent $q$, so that as the universe age goes to zero (the early hot universe), $q\rightarrow0$, where the critical temperature diverges. 
The other point of view is based on a presumable duality between the free $q$-boson system and the interacting ordinary bosons, in which $q$ controls the interactions. If true, then one expects that the interaction between ordinary bosons (corresponding to the $q$ parameter in the dual $q$-boson system) depends on the universe age, and consequently is time dependent. In this case the latter scenario has the benefit of giving us this effective time dependent $q$, which has not been analyzed before. This also predicts that the BEC phase of $q$ bosons is conservative, and no complementary theory is needed to explain the entropy.

In the following we find the temperature dependence of the ground state occupation number of $q$ bosons, and show that for sufficiently small $q$s, all particles settle down in the ground state irrespective to the system temperature. The number of ground state particles is given by:
\bea
 N_{0}(T)=N-N_{e}(T)
\label{i1}
\eea
where $N_{0}$ and $N_{e}$ are the number of particles in the ground and excited states respectively.
Also, the number of particles in the excited states in the condensate phase is
\bea
 N_{e}(T)
=\frac{V}{(q-q^{-1})}(\frac{2\pi mk_{B}T}{h^{2}})^{3/2}{H}_{5/2}( q^{2},1,q), ~~~T\le T^{q}_{c},\nonumber\\
\label{i2}
\eea
while at $T=T^{q}_{c}$, all particles are in excited states, so that:
\bea
N=N_{e}(T=T^{q}_{c})=\frac{V}{(q-q^{-1})}(\frac{2\pi mk_{B}T^{q}_{c}}{h^{2}})^{3/2}{H}_{5/2}( q^{2},k,q).\nonumber\\
\eea
Therefore,  it is straightforward to obtain that\cite{pathria1996statistical}
\bea
 \frac{N_{0}}{N}=1-\frac{N_{e}}{N}=1-\left(\frac{T}{T_{c}^{q}}\right)^{3/2}.
\label{i5}
\eea
But we showed that the critical condensation temperature goes to infinity for small values of $q$. Therefore, ${T}/{T_{c}^{q}}$ vanishes at any finite temperature and consequently $N_{0}/{N}$ goes to unity, meaning that all $q$-bosons with sufficiently small $q$s, live in condensate phase at any finite temperature and occupy the ground state of the system. In such circumstances, the system has only one microstate and the entropy of the system will be zero.
{
\section{Relic density and Mass Bounds}\label{sec5}}
{
Mass bounds for dark matter candidates has been already obtained for particles that decouple in or out of equilibrium for various distribution functions \cite{boyanovsky1999nonequilibrium}. Also, it has been shown that depending on the relation between the critical condensation temperature; $T_c$, and decoupling temperature; $T_d$, a Bose Einsein condensate light relic could act as cold dark matter but the decoupling scale must be higher than the electroweak scale. \cite{boyanovsky1999nonequilibrium}}

{
 Also in \cite{galli2009cmb,goodman2010constraints,boyanovsky2008constraints} some constraints on the dark matter have been extracted using CMB, various colliders such as Tevatron and LHC, galaxy observations, $N$-body simulations and theoretical arguments.
We will extract the mass bounds on the dark matter particles obeying $q$-boson statistics and condensed to ground state.
Let us first take the Friedmann-Robertson-Walker space-time \cite{weinberg2008cosmology} in which the universe is spatially flat with time dependent scale factor $a(t)$, and the dispersion relation is  $\epsilon=\sqrt{m^2+ p_{c}^{2}}$, where $p_c$ denotes the momentum in comoving frame. If a $q$-boson with mass $m$ is put in local thermodynamic equilibrium (LTE) in the early plasma with decoupling temperature $T_d$, its mean occupation number or distribution function reads
\bea
f_{d}(p_{c}) =\frac{1}{q-q^{-1}}\ln\left(\frac{e^{\frac{\sqrt{m^{2}+p_{c}^{2}}-\mu_{d}}{T_{d}}}-q^{-1}}{e^{\frac{\sqrt{m^{2}+p_{c}^{2}}-\mu_{d}}{T_{d}}}-q}\right),
\label{da}
\eea
where, $\mu_d$ is the chemical potential at decoupling. Using kinetic energy-momentum tensor, the energy density and the pressure are given by
\bea
\rho = g\int \frac{d^3P_f}{(2\pi)^3}
\sqrt{m^2+P^2_f}  f_d[a(t) \; P_f],
\label{db}
\eea
\bea
{\cal{P}} =\frac{g}{3}   \int \frac{d^3P_f}{(2\pi)^3}    \frac{
P^2_f}{\sqrt{m^2+P^2_f}}  \; f_d[a(t) \; P_f] ,
\label{dc}
\eea
where, $P_f =p_c /a(t)$ is physical momentum, redshifting with the expansion and $g$ is the number of internal degrees of freedom.
The number of $q$-bosons per comoving volume $V_c$ in thermodynamic limit is evaluated as follows
\bea
n= n_0 + \int\frac{d^3p_c}{(2\pi)^3}\frac{1}{q-q^{-1}}\ln\left(\frac{e^{\frac{\sqrt{m^2+p_c^2}-\mu_{d}}{T_{d}}}-q^{-1}}{e^{\frac{\sqrt{m^2+p_c^2}-\mu_{d}}{T_{d}}}-q}\right),~~~~
\label{dg}
\eea
where
\bea
 n_0=\frac{1}{V_c}\frac{1}{q-q^{-1}}\ln\left(\frac{e^{\frac{m-\mu_{d}}{T_{d}}}-q^{-1}}{e^{\frac{m-\mu_{d}}{T_{d}}}-q}\right).
 \label{dh}
\eea
In the ultra relativistic limit, $ m/T_d \ll1 $, one finds that
\bea
n= n_0 + \frac{T^3_d \;{H}_{4}( q^{2},1,q)}{\pi^2(q-q^{-1})}.
\label{di}
\eea
We notice that for small values of $q$, the condensation temperature is sufficiently higher than the decoupling temperature and the system is in condensate phase where $z_c =q^2$.In fact for arbitrary values of decoupling temperature we can extract that
\bea
1-\frac{n_0}{n} = \Bigg\{\begin{array}{c}
          \Big(\frac{T_d}{T_{c}} \Big)^3 \quad {\rm for} \quad T_d < T_{c} ,\\
          1 \quad {\rm for} \quad T_d>T_{c},
        \end{array}
 \label{dj}
\eea
where the critical temperature is given by
\bea
T_{c} = \Big[ \frac{(q-q^{-1})\pi^2 \;  n}{{H}_{4}( q^{2},1,q)}\Big]^{\frac{1}{3}}\,.
\label{dk}
\eea
However, for small values of $q$, the system always is in the condensate phase independent of the value of the decoupling temperature. Therefore, in the thermodynamic limit the distribution function for $q$-bosons that decouple in ultra-relativistic regime; $ m/T_d\ll 1 $,  is given by
\bea
f_d(p_c) = n_0 \, \delta^{(3)}(\vec{p}_c)+\frac{1}{q-q^{-1}}\ln\left(\frac{e^{\frac{p_c-\mu_{d}}{T_{d}}}-q^{-1}}{e^{\frac{p_c-\mu_{d}}{T_{d}}}-q}\right).~~~~
\label{dl}
\eea
The number of particles per unite physical volume can be evaluated as follows
\bea
n(t)=g\int\frac{d^3 P_f}{(2\pi)^3}f_{d}\left(a(t)P_f\right),
\eea
and using Eq. (\ref{dl}) we obtain that
\bea
 n(t)= n_0(t) +
\frac{{H}_{4}( q^{2},1,q)}{\pi^2(q-q^{-1})} \; T^3_d(t)  ,
 \label{dn}
\eea
where, $T_d(t) = T_{d}/a(t)$ and $n_0(t)=n_0 /a^3(t)$. Using Eqs. (\ref{dj}) and (\ref{dn}), we work out
\bea
n(t) =\frac{{H}_{4}( q^{2},1,q)}{\pi^2(q-q^{-1})} \left(\frac{T_c}{T_d}\right)^3 \; T^3_d(t) \; .
\label{dq}
\eea
Also, using Eqs. (\ref{db}), (\ref{dc}) and (\ref{dl}) we evaluate the energy density and pressure as follows
\bea
\rho(t) & = &  g~m
\left\{ n_0(t) + \frac{1}{(q-q^{-1})}T^3_d(t) \; I^{nc}_\rho[x(t)] \right\},
\label{dr}
\eea
\bea
{\cal{P}}(t) & = &   g~ \frac{1}{(q-q^{-1})}\frac{T^5_d(t) }{3 \, m} \,
I^{nc}_{\cal{P}}[x(t)]
\label{ds}
\eea
where,
\bea
&I^{nc}_\rho[x(t)]  =   \frac{1}{2 \, \pi^2} \int_0^\infty
\sqrt{1+\frac{y^2}{x^2(t)}}\ln\left(\frac{e^{y-\ln{q}^2}-q^{-1}}{e^{y-\ln{q}^2}-q}\right)dy,~~~~~\\
&I^{nc}_{\cal{P}}[x(t)]=
\frac{1}{2 \, \pi^2}\int^\infty_0 \frac{y^4}{\sqrt{1+ \frac{y^2}{x^2(t)}}}
\ln\left(\frac{e^{y-\ln{q}^2}-q^{-1}}{e^{y-\ln{q}^2}-q}\right) dy,~~~~~~~
\label{du}
\eea
where, $x(t) = {m}/{T_d(t)}$ and we notice that for small values of $q$, the fugacity always takes the value $z=q^2$, because the system is in the condensate phase.  Indeed, taking the limit $q\to0$, one finds that the functions $I^{nc}_\rho[x(t)]$ and $I^{nc}_{\cal{P}}[x(t)]$ tend to zero, showing that the main contribution to the energy density comes from the condensate phase and pressure is vanished for small values of $q$. We then find that
\bea
\rho(t) &=& g~ m~ n_{0}(t)\thickapprox g~ m~ n(t)\nonumber\\
&=&g~m~ \frac{{H}_{4}( q^{2},1,q)}{\pi^2(q-q^{-1})}\left(\frac{T_c}{T_d}\right)^{3} T_{d}^{3}(t)
\label{dz}
\eea
For $a$-species particles, the relic abundance \emph{today} has been evaluated as follows \cite{boyanovsky2008constraints}
\bea
\Omega_a \; h^2 = \frac{m_a}{25.67~\mathrm{eV}} \;  \frac{g_a \;
\int^\infty_0 y^2  \; f_{d,a}(y) \; dy}{2 \;  g_{d,a} \; \zeta(3)} ,
 \label{rdb}
\eea
where we used that today $ h^2 \; n_\gamma/\rho_c = 1/25.67 $eV. $n_{\gamma}$ denotes the density of photons and $\rho_{c}$ is the critical density.
If this decoupled species contributes a fraction $ \nu_a $ to dark matter,
with $\Omega_{a}=\nu_{a}\Omega_{DM} $ and using that $\Omega_{DM}h^2 =0.105$ for non-baryonic dark matter, then one finds:
\bea
\nu_a= \frac{m_a}{2.695~\mathrm{eV}} \; \frac{g_a\int^\infty_0 y^2 \;
f_{d,a}(y) \; dy}{2  \; g_{d,a} \; \zeta(3)} \; .
\label{rdc}
\eea
Using this, and knowing that $ 0\leq \nu_a \leq 1 $ we deduce the  constraint
\bea
m_a \leq 2.695 \; \mathrm{eV} \; \frac{2  \; g_{d,a} \; \zeta(3)}{g_{a} \;
\int^\infty_0 y^2  \; f_{d,a}(y) \; dy} \; .
\label{rdd}
\eea
Thus, using the distribution function of deformed bosons in condensate phase, we work out the upper bound for the mass.
For sufficiently small values of $q$, $T_d<T_c $ and using Eq.(\ref{rdb}) we obtain for $q$-deformed BEC
\bea
\Omega_{q-\mathrm{BEC}} \; h^2 =  \frac{m}{25.67~\mathrm{eV}} \;
\frac{g}{g_d}\frac{{H}_{4}( q^{2},1,q)}{(q-q^{-1})\zeta(3)} \left(\frac{T_c}{T_d}\right)^3 .
\label{rde}
\eea
Therefore, the $q$-BEC dark matter fraction which contribute is given by
\bea
\nu_{q-\mathrm{BEC}} = \frac{m }{2.695~\mathrm{eV}} \; \frac{g}{g_d}\frac{{H}_{4}( q^{2},1,q)}{(q-q^{-1})\zeta(3)}
\left(\frac{T_c}{T_d}\right)^3,
\label{rdf}
\eea
and consequently the {upper} bound is evaluates as follows
\bea
m  \leq  2.695   \;  \frac{g_d}{g}\frac{(q-q^{-1})\zeta(3)}{{H}_{4}( q^{2},1,q)} \;
\left(\frac{T_d}{T_c}\right)^3 ~  \mathrm{eV} \; .
\label{boundq}
\eea
The velocity dispersion when the particle becomes non-relativistic is evaluated as follows
\bea
\Big\langle \vec{V}^2   \Big\rangle  =  \Big\langle \frac{\vec{P}^2_f}{m^2} \Big\rangle =  12(\frac{{H}_{6}( q^{2},1,q)}{{H}_{4}( q^{2},1,q)})\Bigg[\frac{T_d(t)}{m} \Bigg]^2\Big(\frac{T_d}{T_c}\Big)^3
\label{rdg}
\eea
Let us now consider the coarse grained (dimensionless) primordial phase space density which is defined by \cite{boyanovsky2008constraints}
\bea
\mathcal{D}  \equiv \frac{n(t)}{\big\langle {\vec{P}^2}_{f}
\big\rangle^\frac32} \; ,
 \label{rdh}
\eea
which is Liouville invariant and $ \big\langle \vec{P}^2_f \big\rangle $ is defined by Eq.(\ref{rdg}). We obtain $\mathcal{D}$ for $q$-BEC as follows
\bea
 \displaystyle \mathcal{D} = \frac{ g}{12^{\frac{3}{2}}\pi^2(q-q^{-1})}\frac{({H}_{4}( q^{2},1,q))^{\frac{5}{2}}}{({H}_{4}( q^{2},1,q))^{\frac{3}{2}}}
(\frac{T_{c}}{T_{d}})^{\frac{15}{2}}.
\label{rdi}
\eea
In the non-relativistic limit,  $ \rho(t) = mn(t) $
and $ \big\langle \vec{V}^2 \big\rangle = \big\langle\frac{\vec{P}^2_f}{m^2} \big\rangle $. We obtain that
\bea
\mathcal{D} = \frac{\rho}{m^4\,\big\langle \vec{V}^2 \big\rangle^{\frac{3}{2}} } =
\frac{Q_{DH}}{m^4},
\label{rdj}
\eea
where, $ Q_{DH} $ is the phase space density introduced by Dalcanton and Hogan \cite{dalcanton2001halo,hogan2000new} and is given by
\bea
Q_{DH}= \frac{\rho}{\big\langle \vec{V}^2 \big\rangle^{\frac32}}.
\label{rdk}
\eea
In the nonrelativistic regime $ \mathcal{D} $ is related to another coarse grained phase space density, $ Q_{TG} $ which is called Tremaine and Gunn density \cite{tremaine1979dynamical} which is defined as follows:
\bea
Q_{TG}= \frac{\rho}{m^4 \; (2 \, \pi \; \sigma^2)^\frac32} =  \left( \frac3{2 \, \pi}
\right)^\frac32 \; \mathcal{D} \; .
\label{rdl}
\eea
The phase space density $, \rho/\sigma^3 $ is the observationally accessible quantity, where $\sigma=\sqrt{{\big\langle \vec{V}^2 \big\rangle/3}}$ is the one dimensional velocity dispersion.
Using $ \rho = m n $  for decoupled particles that are non-relativistic today, the primordial phase space density is defined as follows
\bea
\frac{\rho_{DM}}{\sigma^3_{DM}} = 3^\frac32   m^4  \mathcal{D}
\equiv 6.611 \times 10^8   \mathcal{D}
\Big[\frac{m}{\mathrm{keV}}\Big]^4
\frac{M_\odot/\mathrm{kpc}^3}{\big(\mathrm{km}/\mathrm{s}\big)^3}
\label{rdm}
\eea
where we notice that $ \mathrm{keV}^4 \; \big(\mathrm{km}/\mathrm{s}\big)^3 =1.2723 \; 10^8 \; \frac{M_\odot}{\mathrm{kpc}^3} $.}

{The phase mixing increases the density and velocity dispersion during collisionless gravitational dynamics so that the coarse grained phase space density remains constant or decreases. Therefore,
\bea
\frac{\rho}{\sigma^3} \leq 6.611 \times
10^8  \;  \mathcal{D} \;  \Big[\frac{m}{\mathrm{keV}}\Big]^4  \;
\frac{M_\odot/\mathrm{kpc}^3}{\big(\mathrm{km}/\mathrm{s}\big)^3}  \;
\,.
\label{rdn}
\eea
We can also evaluate $ \mathcal{D} $ in the ultrarelativistic limit for the $q$ deformed particles that decouple from the early plasma. It does not depend on
the mass. Using Eq. (\ref{rdn}) one can find a {lower} bound on the mass of the $q$-BEC particles directly from the observed phase space density and
the distribution function of deformed bosons in condensate phase.}

{
New bounds have been extracted from the latest compilation presented directly on $ \rho/\sigma^3 $ for the data set
comprising ten satellite galaxies in the Milky-Way dSphs \cite{boyanovsky2008constraints,wyse2007observed}. First we rewrite the Eq.(\ref{rdn}) as follows
\bea
m^4 \geq \frac{\Big[62.36~\mathrm{eV}\Big]^4}{\mathcal{D}} \; 10^{-4} \;
\frac{\rho}{\sigma^3} \; \frac{\big(\mathrm{km}/\mathrm{s}\big)^3}{M_\odot/\mathrm{kpc}^3}
\; ,
\label{rdo}
\eea
The data of Ref. \cite{wyse2007observed} indicates that
\bea
0.9 \leq 10^{-4} \; \frac{\rho}{\sigma^3} \;
\frac{\big(\mathrm{km}/\mathrm{s}\big)^3}{M_\odot/\mathrm{kpc}^3}
\leq 20 \; ,
\label{rdp}
\eea
and for a value in the middle of the above range, the lower bounds reads
\bea
m \gtrsim \frac{100}{\mathcal{D}^{\frac{1}{4}}} ~\mathrm{eV} \; .
\label{rdq}
\eea
Using Eqs. (\ref{rdi}) and (\ref{rdq}), for $q$-bosons in condensate phase ($T_d<T_c$) we work out the lower bound on the mass of particles as follows
\bea
m \gtrsim  \frac{100\times 12^{\frac{3}{8}} \pi^{\frac{1}{2}}(q-q^{-1})^{\frac{1}{4}}}{g^{\frac{1}{4}}}\frac{({H}_{6}( q^{2},1,q))^{\frac{3}{8}}}{({H}_{4}( q^{2},1,q))^{\frac{5}{8}}}\Bigg[\frac{T_d}{T_c}\Bigg]^{\frac{15}{8}}.
\label{rdr}
\eea}

{
We notice that if the distribution function is known, $\mathcal{D}$ is evaluated by Eq.(\ref{rdi}) completely at decoupling. Therefore, the lower an upper bounds of mass will be calculated for particles that decouple out of local thermodynamics equilibrium which is consistent with observational data.
Combining the {upper} bound Eq. (\ref{boundq}) with the {lower}
bound Eq. (\ref{rdo}), we establish the mass range of $q$ bosons condensate as dark matter candidate
\bea
\frac{416.06 \mathrm{eV}(q-q^{-1})^{\frac{1}{4}}(H_{6}(q^{2},1,q))^{\frac{3}{8}}}{g^{\frac{1}{4}}(H_{4}(q^{2},1,q))^{\frac{5}{8}}}\left(\frac{T_{d}}{T_{c}^{q}}\right)^{\frac{15}{8}}
\leq\nonumber\\m^{q}\leq \frac{2.695 g_{d}(q-q^{-1})\zeta(3)}{g H_{4}(q^{2},1,q)}\left(\frac{T_{d}}{T_{c}^{q}}\right)^{3} \mathrm{eV}.
\label{rdt}
\eea
The same relation for the mass bounds of ordinary boson condensate has been extracted as follows \cite{boyanovsky2008constraints}
\bea
\frac{360 \mathrm{eV}}{g^{\frac{1}{4}}}\left(\frac{T_d}{T_{c}^{b}}\right)^{\frac{15}{8}}\le m^b \le \frac{g_d}{g} 2.695 \left(\frac{T_d}{T_{c}^{b}}\right)^{3}\mathrm{eV},
\label{rdbb}
\eea
where $ g_d $ is the number of ultrarelativistic degrees of freedom at decoupling. If we suppose that the decoupling temperature and the ultrarelativistic degrees of freedom for both BEC and $q$-BEC are the same, using the related condensation temperature, we can argue that
\bea
0.97\frac{({H}_{6}( q^{2},1,q))^{\frac{3}{8}}}{(q-q^{-1})^{\frac{3}{8}}}
\leq \frac{m^{q}}{m^{b}}
\leq 1.
\label{rdu}
\eea
Therefore, the upper bound of mass for ordinary and deformed bosons condensate as a model of dark matter is the same while the lower bound of $q$-BEC depends on the value of $q$. In previous sections we argued that the condensation temperature of deformed boson gas tends to infinity and the system always lives in condensate phase for small values of $q$. Fig. (\ref{fig5}) shows that in this limit, the lower bound tends o zero. Therefore, $q$-BEC as a model of dark matter is consistent with light non interacting dark matter.
\begin{figure}
\includegraphics[scale=0.45]{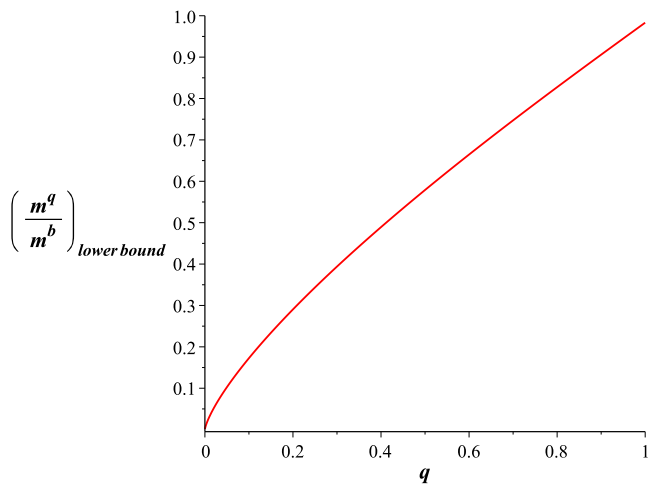}
\caption{The lower bound of mass for $q$ deformed bosons in condensate phase. For sufficiently small values of $q$, the lower bound tends to zero.}
\label{fig5}
\end{figure}}

{ For BEC model, it is obvious from Eq. (\ref{rdbb}) that for $T_{d}\ll T_{c}$, the fulfillment of the bound
requires very large $g_d$. Namely, in the presence of a BEC thermal decoupling occurs at a scale much larger than the
electroweak scale for $T_{d}\ll T_{c}$ \cite{boyanovsky2008constraints}. However for $q$-BEC model, for small values of $q$, naturally $T_{d}\ll T_{c}$ is established. Using Eqs. (\ref{rdt}) and (\ref{rdu}), we demonstrate that the existence of a $q$-dependent factor which tends to zero in the limit of small $q$ in lower bound, relaxes the large $g_d$ condition, meaning that $g_d$ is not needed to be large since the inequalities are satisfied for small $q$ values.}\\

\section{Conclusion}\label{sec6}
We considered the condensate of an ideal $q$-boson gas as a model of dark matter. We argue that because of some reasons, $q$-boson condensate with enough small value of deformation parameter is compatible with the properties of dark matter.

It has been shown that the BEC as a model of dark matter gives some useful information about the density profile, rotational velocity and mass profile. In fact a good agreement between theoretically BEC model predicted rotation curves and the observational data. { However, we expect that the temperature of universe to be high in past epochs. If light bosons are produced by thermal processes, then it will result in a too large number of relativistic degrees of freedom. Thus, the temperature of the dark matter candidate is not equal to the age of the Universe. We considered that the $q$-boson condensate could be a valid an compatible model for all epochs.} We obtained the critical condensation temperature of $q$-bosons and found that for small values of deformation parameter, the critical temperature goes to infinity. Therefore, the system lives in condensate phase at any finite temperature.

The equation of stat of an ideal $q$-boson gas at condensate phase returns to the equation of state on an ideal classical gas for small value of deformation parameter. Considering $q$-boson condensate as a model of dark matter is in consistency with the non interacting nature of dark matter.

We evaluated the entropy of $q$-bosons at condensate phase and showed that it is vanished at in the small value limit of deformation parameter. The entropy of ordinary bosons condensate depends on the temperature and by decreasing the temperature, the entropy will be decreased.
In order to compensate of this reduction of entropy, the system have to interact with surroundings or remnant of the galaxy. This is in contrast with the non interacting behaviour of dark matter. Zero entropy of $q$-boson condensate for small values of $q$ at any finite temperature indicates that the proposed model is consistent with the non interacting behaviour of dark matter.

We argued that all $q$-bosons with small value of deformation parameter, occupy the ground state at any finite temperature. In fact, the particles never has been excited and live in ground state for all time duration. Maybe, non detectability of constituents of dark matter is related to the non excitability of such particles.

{Also, we investigated the mass bounds on the condensate of $q$ deformed bosons as a model of dark matter. Using the phase space density which is evaluated by distribution function of condensed $q$ bosons and observational date, the lower and upper bounds of mass were obtained. We showed that the upper bound of $q$-BEC and ordinary BEC model is the same, while for small values of $q$, the lower bound tends to zero which is in favor of the light dark matters particles.}


\bibliography{re}

\end{document}